\documentclass[fleqn,10pt]{wlscirep}
\usepackage[utf8]{inputenc}
\usepackage[T1]{fontenc}

\usepackage{cite}
\usepackage{amsmath,amssymb,amsfonts}
\usepackage{algorithmic}
\usepackage{graphicx}
\usepackage{textcomp}
\usepackage{xcolor}
\usepackage{xspace}
\usepackage{comment}
\usepackage{cancel}
\usepackage{graphicx}
\usepackage{subcaption}
\usepackage[hang,flushmargin]{footmisc}
\usepackage[font={bf,it,footnotesize},labelsep=colon,justification=centering]{caption}
\iftrue
\newcommand{\chaterji}[1]{\textcolor{blue}{SC: #1}}
\newcommand{\suryavansh}[1]{\textcolor{red}{SS: #1}}
\else
\newcommand{\chaterji}[1]{}
\newcommand{\suryavansh}[1]{}
\fi

\newcommand{\name}{{Ambrosia}\xspace}

\UseRawInputEncoding
\usepackage{array}
\newcolumntype{P}[1]{>{\centering\arraybackslash}p{#1}}
\newcolumntype{M}[1]{>{\centering\arraybackslash}m{#1}}

\usepackage{hyperref}
\usepackage[numbers]{natbib}
\usepackage{multirow}
\usepackage{dblfloatfix}

\iftrue
\newcommand{\somali}[1]{\todo[inline,color=red!40]{somali: #1}}
\newcommand{\shikhar}[1]{\todo[inline,color=blue!40]{shikhar: #1}}
\else
\newcommand{\somali}[1]{}
\newcommand{\shikhar}[1]{}
\fi

\begin{document}
\pagestyle{plain}

\title{\name: Reduction in Data Transfer from Sensor to Server for Increased Lifetime of IoT Sensor Nodes}

\author[1]{Shikhar Suryavansh}
\author[2]{Abu Benna}
\author[3]{Chris Guest}
\author[4, *]{Somali Chaterji}

\affil[1]{Cisco Systems, United States}
\affil[2, 3]{Beaconchain, Canada}
\affil[4]{Purdue University, West Lafayette, IN}

\affil[*]{schaterji@purdue.edu}




\begin{abstract}
Data transmission accounts for significant energy consumption in wireless sensor networks where streaming data is generated by the sensors. This impedes their use in many settings, including livestock monitoring over large pastures (which forms our target application). We present \name, a lightweight protocol that utilizes a window-based timeseries forecasting mechanism for data reduction. \name employs a configurable error threshold to ensure that the accuracy of end applications is unaffected by the data transfer reduction. Experimental evaluations using LoRa and BLE on a real livestock monitoring deployment demonstrate 60\% reduction in data transmission and a 2X increase in battery lifetime.

\end{abstract}

\maketitle

\section{Introduction}
\label{sec:introduction}

There has been a tremendous growth in the number of IoT devices deployed worldwide. These IoT devices comprise of a network of dedicated physical objects (things) that contain embedded technology to interact with the external environment. According to Cisco, 
the number of networked devices is expected to reach 27.1 billion in 2021, i.e., about 3.5 devices per human on the planet. These IoT wireless nodes typically consist of a microcontroller, transceiver, memory unit, power source, and one or more sensors for sensing the ambient environment. Examples of such sensors include accelerometers, and sensors for vibration, temperature, and humidity. Data collected by the sensors is either processed locally or sent to a server for analysis. This is meant to support algorithmic processing of sensor data for varied use cases like anomaly detection and object detection for end user applications in home automation, digital agriculture, smart factories, etc. 

Since the sensor nodes are constrained in terms of their computational capacity and energy budget, the processing of the collected data usually takes place at a nearby connected edge server. Edge servers can include devices such as Raspberry Pis, wireless routers, or low to mid-range servers installed close to the sensor nodes to reduce latency in data transmission. Low latency is important because the data may be used to trigger real-time responses, in addition to monitoring purposes, such as activating cooling procedures if the temperature of an industrial plant operation rises above a threshold.

Several wireless network technologies can be utilized for data transfer between the sensor nodes and the edge server. These include Bluetooth Low Energy (BLE), IEEE 802.11 power saving mode, IEEE 802.15.4/e, as well as long-range technologies such as LoRa and SIGFOX. Low energy consumption is a critical requirement for IoT sensor nodes since they are usually battery operated. In order to increase battery life, energy consumption of these sensor nodes have been optimized using techniques such as efficient routing~\cite{efficient_routing}, data compression~\cite{dias2016survey}, duty cycling~\cite{duty_cycling_1}, mobility~\cite{tirta2004efficient}, and approximate computing~\cite{venkataramani2015approximate}. 

Most of the energy consumption at the sensor nodes occurs because of data transmissions. Irrespective of the wireless network technology used, data transfer accounts for at least $85\%$ of the total energy consumption~\cite{zhao2019understanding}. 
A comparison of the energy consumption of various network technologies obtained from~\cite{comparison_of_device_lifetime} has been provided in Table~\ref{table:power_consumption}. The percentage is much higher for long-range technologies such as LoRa and SIGFOX, where data transmission can consume as high as $99\%$ of the available energy. Therefore, reduction in data transmission from the sensor nodes to the edge nodes or to the cloud (as the case may be for the processing requirement) can significantly reduce the energy consumption and increase battery lifetimes.

\begin{table*}
\centering
\renewcommand{\arraystretch}{1.5}
\begin{tabular}{ |M{2cm}|M{2cm}|M{2cm}|M{2cm}|M{2cm}|M{2cm}|M{2.1cm}|}
 \hline
 \multicolumn{7}{|c|}{Power Consumption Comparison of Wireless Network Technologies} \\
 \hline
 Wireless Technology & Hardware Platform & P\_Tx (mW) & P\_Rx (mW) & P\_Idle (mW) & P\_Sleep ($\mu$W) & Percentage Power Consumption for Data Transfer \\[5 pt]
 \hline
 \textbf{802.11 PSM} & G2M5477 & 699.6  & 170 & 66 & 13.2 & 92.9 \% \\
 \hline 
 \textbf{BLE} & nRF51822 & 37.2 & 42.3 & 13.2 & 7.8 & 85.8 \% \\
 \hline
 \textbf{802.15.4} & SmartMeshIP & 24.11 & 20.87 & 4.67 & 4.32 & 90.6 \% \\
 \hline
 \textbf{LoRa} & GreenNet & 419.6 & 44.06 & --- & 4.32 & 99.9 \% \\
 \hline
 \textbf{SIGFOX} & GreenNet & 147 & 39 & --- & 4.32 & 99.9 \% \\
 \hline
\end{tabular}
\caption{\textbf{Power consumption of various network technologies} (assuming all the states are equally likely; for high data rate, the likelihood of sensor node being in Tx or Rx would be greater than the other states resulting in an even higher percentage power consumption for data transfer)}
\label{table:power_consumption}
\vspace{-4 mm}
\renewcommand{\arraystretch}{1}
\end{table*}

In this paper, we propose \name, an efficient technique for reducing the data transmission from the sensor nodes to the servers\footnote{In ancient Greek mythology, \name is the food and drink of the Greek gods, which brings longevity to whoever consumes it. In our work, we aim at increasing the lifetime of the IoT sensor nodes by reducing the bulk of  data transfer.}. {\em The key intuition behind \name is that data samples that can be successfully predicted at the edge server do not need to be transmitted from the sensor nodes.} \name utilizes a simple window-based prediction scheme to decide which data samples need to be transmitted from the sensor node to the server. Since the sensor nodes have limited resources, a computationally intensive technique for data reduction, no matter how accurate, would be sub-optimal. Executing a compute-intensive technique on the sensor node would overshadow the advantage accrued from data reduction. Further, a computationally intensive algorithm may not be able to keep pace with the rate at which streaming data is being generated resulting in missing timing deadlines. 
Thus, we design \name to be sufficiently lightweight to be able to easily execute on the constrained sensor nodes, as well as designed toward reduced data transmission to the edge servers while meeting the user-defined accuracy bounds. 

The impact of errors in the transmitted data on the accuracy is different for different types of applications. Hence, the optimal value of the error threshold would depend upon the specific application under consideration. For example, a sensitive application such as operating a medical equipment based on the sensor readings would be susceptible to even a small error in the data. On the other hand, applications such as anomaly detection can tolerate a higher magnitude of error in the transmitted data. 
We utilized data collected from different production-grade sensor nodes in a livestock monitoring deployment on a large farm. We measured the effectiveness of \name in different application settings with different tolerances for errors. In particular, for the anomaly detection application, \name is hugely effective in reducing the amount of data transmitted from the sensor nodes to the edge servers.

This problem statement of reducing transmission in sensor networks has seen significant work (as pointed out with categorization above~\cite{efficient_routing, dias2016survey, duty_cycling_1, tirta2004efficient, venkataramani2015approximate}). However, our approach is novel in that it creates a very lightweight approach, one that can run even on the most constrained of sensor nodes, like ear tags on livestock. Further, no prior work has looked at the interplay between short-range and long-range wireless technologies (BLE and LoRa respectively) and has not shown the effect of the error threshold on anomaly detection  --- we use a recent approach from the ML literature for {\em application-agnostic} anomaly detection~\cite{rrcf}.


The main contributions of this paper are:

\begin{enumerate}
    \item We propose \name, a lightweight protocol that can reduce the number of data samples that need to be transmitted from the IoT sensor node to the edge server in a wireless sensor network, thereby reducing energy consumption. We show that \name can reduce the data transmission by at least $60 \%$ thereby increasing lifetime by at least $1.7$X for low traffic intensities and $2$X for high traffic.
    
    \item We take into consideration the variation in the error tolerance of different applications and design our technique in a way that ensures that the accuracy of the applications is not compromised because of reduction in the data transmission. For an error-sensitive application like displacement computation, we obtain $\ge$ $35\%$ reduction in data transfer whereas for an error-tolerant application like anomaly detection, the reduction in data transfer is more significant [$\ge$ $68\%$].

    \item We perform evaluation with multiple sensors (such as MEMS vibration sensors, temperature and humidity sensors) and applications, including a real-world deployment in a production livestock farm, to demonstrate that \name is applicable to a wide range of sensors and applications.
\end{enumerate}

The rest of the paper is organized as follows: Section~\ref{sec:background} presents the background on wireless sensor networks and their energy consumption. Section~\ref{sec:system_model} provides the details of our system model and the proposed protocol. Section~\ref{sec:experimental_setup} presents the experimental setup and Section~\ref{sec:evaluation} the evaluation results. Finally, Section~\ref{sec:conclusion} concludes the paper.

\section{Background and Related Work}
\label{sec:background}

    
   \noindent  \textbf{Power consumption of different wireless technologies}
   
   Different wireless communication technologies can be used for data transmission in an IoT sensor node depending upon the required data rate, data size, and range. We present a comparison of the power consumption of the IoT sensor node using various wireless technologies in Table~\ref{table:power_consumption} obtained from~\cite{comparison_of_device_lifetime}. 
   From Table~\ref{table:power_consumption}, it is evident that majority of the power is consumed for data transmission and reception for all the technologies. Moreover, for long-range technologies such as LoRa and SIGFOX, the amount of power consumed for data transmission is higher than that for low-range technologies such as BLE ( ($99.9\%$ versus $85.8\%$). Thus, \name can be more beneficial in energy savings when deployed for long-range technologies.
   
    \noindent \textbf{Time-series forecasting}
    State space models (SSMs) provide a principled framework for modeling and learning time series patterns, with prominent examples being ARIMA models and exponential smoothing. 
    SSMs predict the future values of data streams indexed by time, based on previously observed values. In a forecasting method without online tuning (offline mode), the predicted values of previous samples are used for future predictions. In contrast, with online tuning, future samples are always predicted using the true values of the previous samples. 
    Online means the model is re-calibrated based on the actual values observed \textit{at runtime}. 
    In the online mode, it is assumed that the true values of the samples would be available after prediction. For our data-reduction protocol, we propose a window-based forecasting technique (Section~\ref{subsection:window_forecasting}) that utilizes a combination of the online and offline modalities. 
    A popular and widely used technique for time-series forecasting is the {\bf AutoRegressive Integrated Moving Average (ARIMA)} model. 
    Time-series forecasting plays an important role in our data-reduction protocol. We provide a comparison of our window-based forecasting protocol vis-\`a-vis ARIMA in Section~\ref{subsection:window_vs_ARIMA}. There are other neural network-based models, e.g., \cite{deepar2020, rangapuram2018deep} that can extract higher-order features and complex patterns within and across time series, but these models are too compute-intensive for our purposes.

\begin{figure}[b]
\vspace{-4mm}
\centering
\includegraphics[width=\columnwidth]{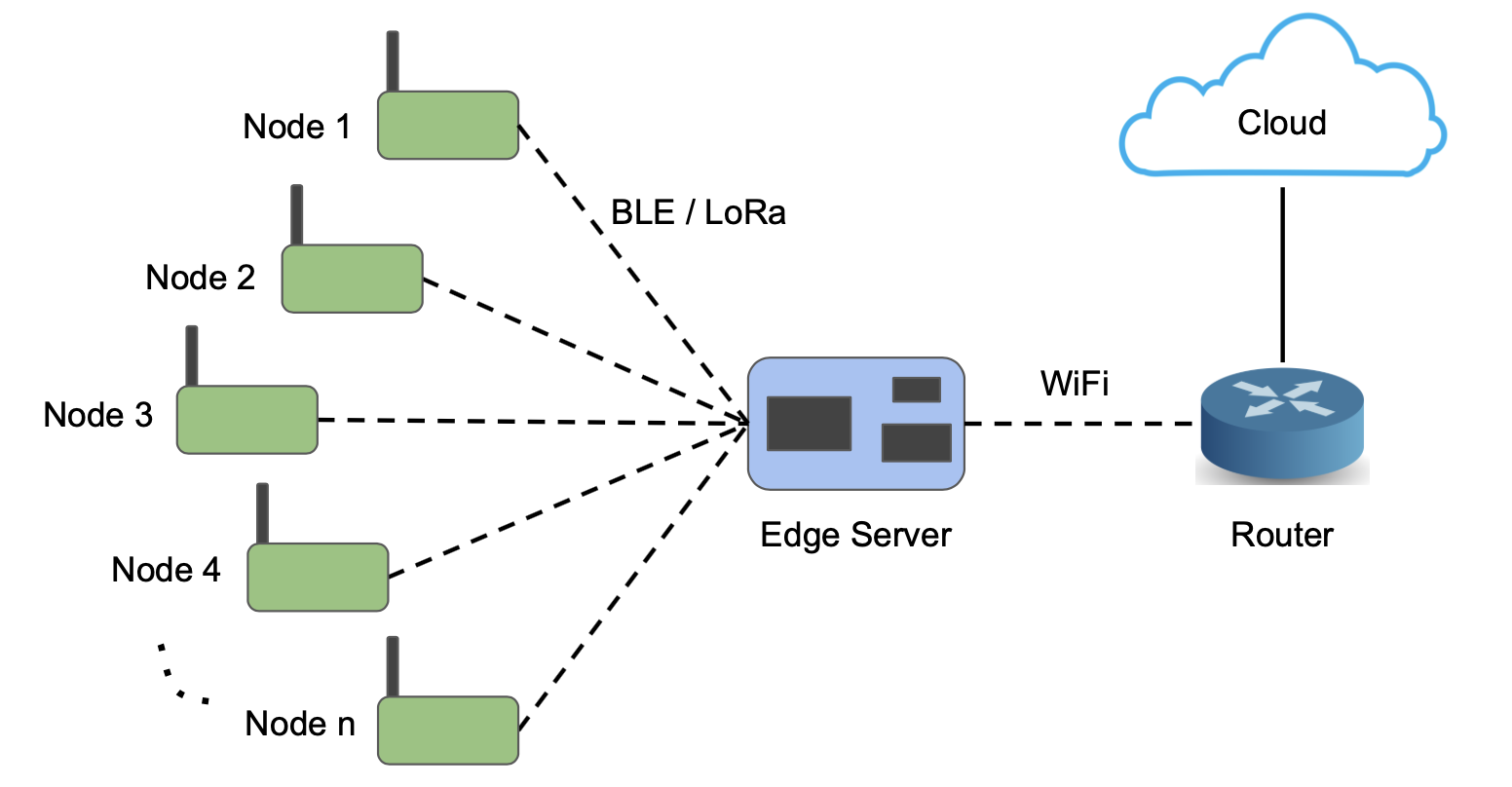}
\vspace{-4mm}
\caption{\label{fig:network_architecture}Network Architecture}
\centering
\vspace{-2mm}
\end{figure}

\begin{figure*}[ht]
\vspace{-4mm}
\centering
\includegraphics[width=0.9\columnwidth]{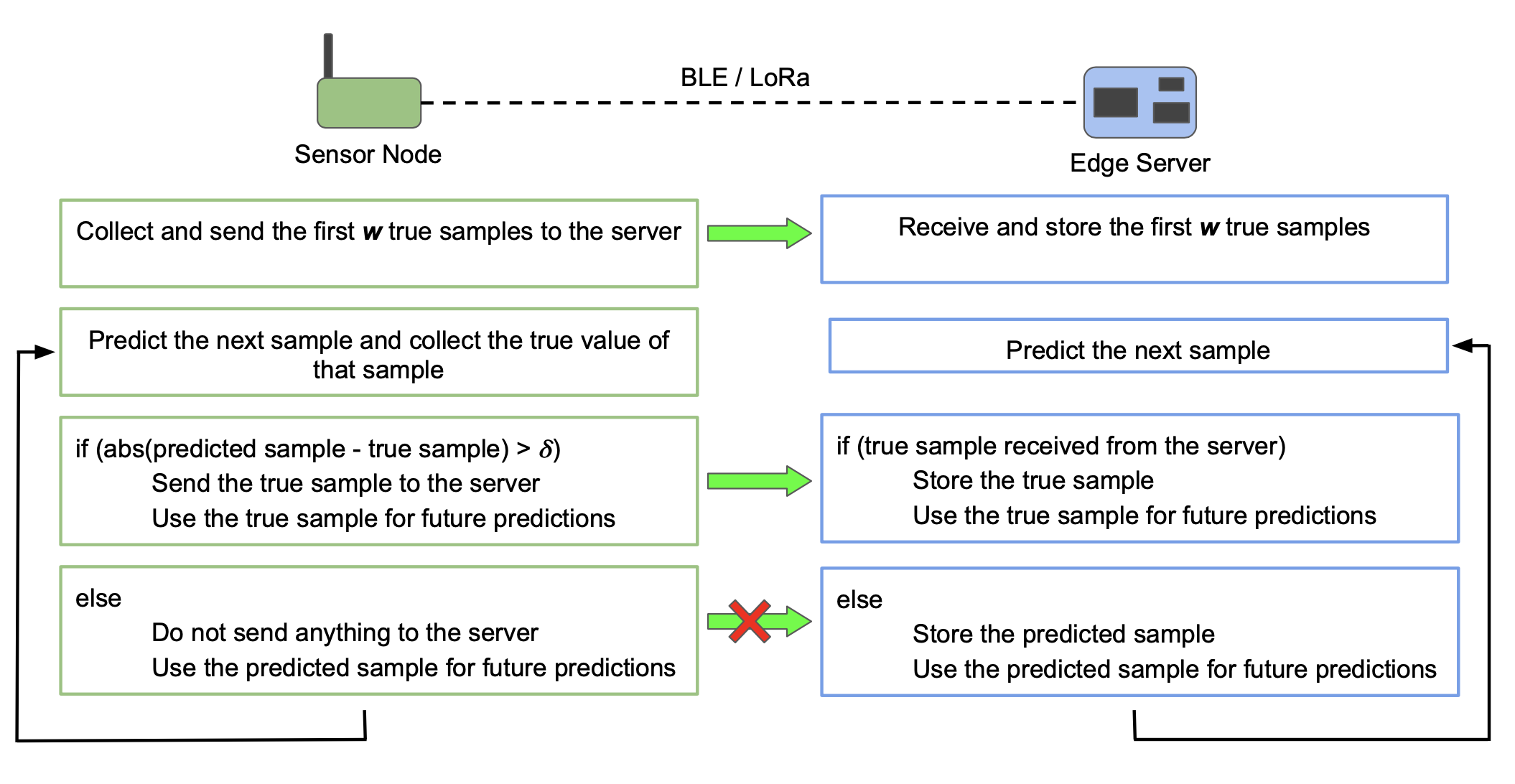}
\vspace{-4mm}
\caption{\label{fig:protocol} Proposed Protocol \name: \textit{\textbf{w}} is the window size and \textit{\textbf{$\delta$}} is the error threshold}
\centering
\vspace{-3mm}
\end{figure*}

    \noindent \textbf{Anomaly detection using Robust Random Cut Forest Algorithm (RRCF)}
    
    RRCF~\cite{rrcf} is a scheme that utilizes an ensemble, robust random-cut data structure, for detecting anomalies from IoT sensor data streams. RRCF does not have a preconceived notion of anomaly. It defines anomalies from the viewpoint of model complexity and determines a point as anomalous if the complexity of the model increases substantially with the inclusion of that point in the data stream. The anomaly score of a data point is obtained using its Collusive Displacement (CoDISP) with the outliers resulting in large CoDISP values. A point is labeled an anomaly if its anomaly score exceeds a threshold. 

\section{System Model}
\label{sec:system_model}

\subsection{Network Architecture}

The network architecture consists of multiple IoT sensor nodes connected wirelessly to an edge server\footnote{Henceforth, when we refer to a {\em ``server''}, we mean an {\em edge server}, rather than a cloud server (unless explicitly mentioned otherwise).} using BLE or LoRa network as shown in Figure~\ref{fig:network_architecture}. 
The edge server is in turn connected to the cloud. Data collected by the IoT sensor nodes is sent to the edge server for analysis. From the edge server, the data can be sent to the cloud for storage and further analysis as needed.

\subsection{Key intuition behind \name}
The main goal of our protocol \name is to reduce the amount of data that is sent from the sensor node to the edge server thereby conserving power and increasing battery life. {\em The animating intuition is that the sensor node does not need to send the data samples whose values can be predicted accurately at the server} --- accurately implies within the bounds of a specified error threshold. If the sensor node can track the sample values that will be predicted successfully at the edge server even if the data sample is not sent, then the sensor node can make an informed decision as to whether or not to send those data samples to the server. This decision would be based on the error between the true sample and the predicted sample value.  

Figure~\ref{fig:protocol} shows the system model for our protocol \name, showing the steps performed at the sensor node and the edge server. The first $w$ (window size) true samples collected by the sensor node are sent to the server. For every sample after that, the sensor node predicts its value and compares the predicted value with the collected true value. The true sample is sent to the server only if the absolute difference between the true and the predicted samples is greater than a user-specified (equivalently, application-specific) error threshold $\delta$. A simple window-based forecasting scheme (described in Section~\ref{subsection:window_forecasting}) is used for prediction. The same prediction scheme is used at the server. This assures that if the true sample is not sent, the value predicted by the server would be within the value generated by the sensor node $\pm$ the error threshold. 

There is another important construct of the protocol needed to assure that there is no mismatch between predicted values at the sensor node and those at the server. When the sensor node decides not to send the true value of a particular sample to the server, it uses the predicted value of that sample for future predictions and {\em not} the true value although it has access to the true value. This is to replicate the settings at the server, namely, the server does not have the true value of that sample and can only use the predicted value for future predictions. 

Hence, the proposed protocol has two important features:

\begin{enumerate}
    \item True samples are sent from the sensor to the server only when required, i.e., when the difference between the true and predicted sample values crosses the error threshold.
    \item The samples for which the true value is not sent from the sensor to the server, the predicted value at the server is the same as the value that was predicted at the sensor (using which it was decided whether to send that sample or not). Hence, when the true sample is not sent, the predicted sample at the server never crosses the error threshold.
\end{enumerate}


       

      \begin{figure*}[t]
        \centering
        \begin{subfigure}[h]{\columnwidth}
            \centering
            \includegraphics[width=0.9\columnwidth]{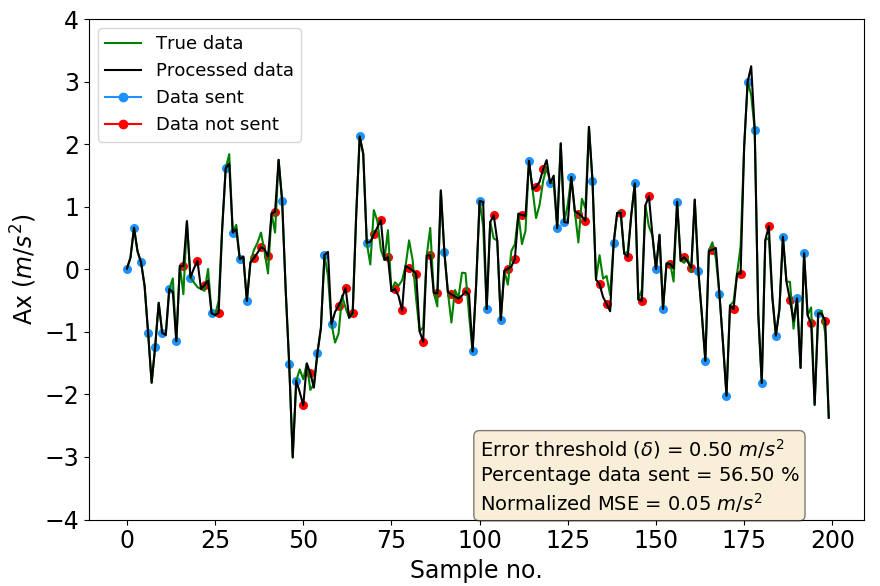}
            \caption[] {Data sent for error threshold = 0.50}    
            \label{fig:prediction_0.5}
        \end{subfigure}
        \hfill
        \begin{subfigure}[h]{\columnwidth}  
            \centering 
            \includegraphics[width=0.9\columnwidth]{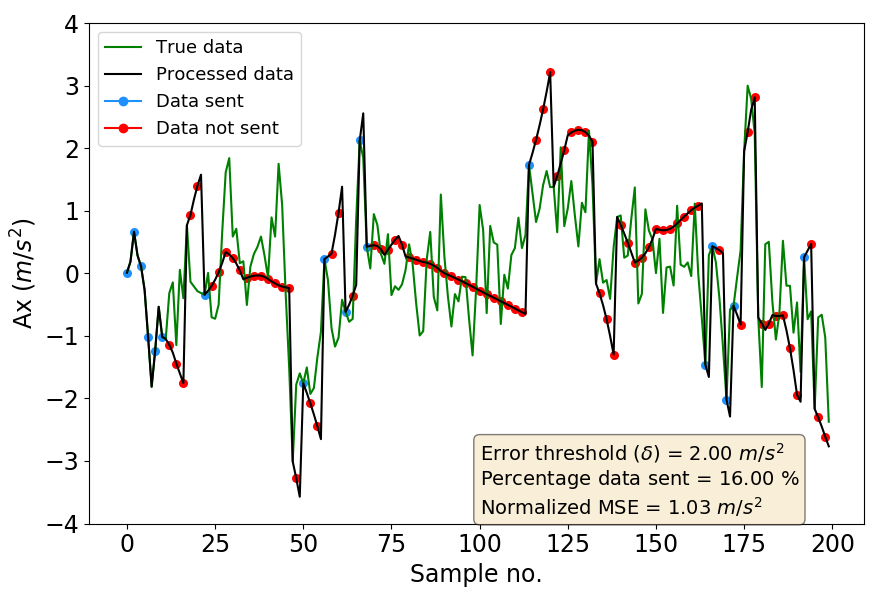}
            \caption[]{Data sent for error threshold = 2.00}
            \label{fig:prediction_2.0}
        \end{subfigure}
        \vskip\baselineskip
        \vspace{-15 pt}
        \caption[]
        {Data sent for various error thresholds. Expectedly, as the threshold becomes higher, fewer data points are sent and the deviation between the true data and the processed data also increases.} 
        \label{fig:true_predicted}
        \vspace{-4 mm}
    \end{figure*}

\subsection{Window-Based Forecasting}
\label{subsection:window_forecasting}

The forecasting technique should be simple enough to run on the energy- and memory-constrained sensor nodes without incurring high costs. It may not be feasible to run a computationally intensive algorithm on the sensor nodes for a slightly improved prediction. This rules out using algorithms like Recurrent Neural Networks (RNN), which have been used for time-series prediction but have also been found to be rather resource intensive~\cite{rangapuram2018deep}. Further, for certain applications, an approximate prediction may be sufficient. We propose a window-based forecasting technique for predicting the next sample using the past samples. The next sample is predicted by adding the average of the difference between $w$ adjacent previous samples to the current sample. The parameter $w$ is the {\em window size}. In Section~\ref{subsection:window_vs_ARIMA}, we compare the performance of this scheme with ARIMA, a traditional approach for time-series forecasting and one which is heavier weight than our calculation. Our window-based forecasting method simplifies as follows:

\vspace{-12 pt}

\begin{gather}
t[n+1]  = t[n] + \frac{1}{w}\sum_{k=1}^{w} t[n-(k-1)] - t[n-k] \\
t[n+1] = t[n] + \frac{1}{w}(t[n] - t[n-w]) = t[n] + d[n] 
\end{gather}

The next sample is predicted using the current sample and $w^{th}$ past sample. A lower value of $w$ implies that a more recent sample is used for the prediction of the next sample. Also, the normalization by $w$ guarantees that if a recent past sample is used, its contribution to the prediction is higher. The optimal value of $w$ would depend upon the IoT data collected and the application.

\section{Experimental Setup}
\label{sec:experimental_setup}

The hardware used in this experiment consisted of 2 parts: Tag nodes attached to the animals, and gateways connected to a PC. 
The tag nodes were standard Lightbug LoRa GPS trackers modified to run custom firmware and encased in 3D printed plastics with ear tag attachment. The electronic component of these devices consisted of a LoRa modem (Semtech SX1726 chipset), pig tail helical antenna, atmel 8-bit microcontroller and MPU6050 6 axis accelerometer.


Once configured, the devices could be set to an ``active'' state. In this mode, they streamed sampled inertial and GPS data to the gateway over LoRa. Quaternions were computed on the device using instantaneous unfiltered data and only sent at 1Hz to reduce data transmission size. Data was buffered locally and then transmitted in batches to the gateway over LoRa at a data rate of approximately 1kB/s, in chunks of 60 bytes (due to module/band licensing restrictions). 



Once the animals were tagged with devices, they were released into a small enclosure to ensure they could be observed and stayed within transmission range. Data was then recorded for activities like walking, eating, or running. In the following section, we present our evaluation results on the acceleration dataset. Similar results were also obtained for other datasets such as temperature, humidity, and vibration [not shown due to space constraints].

\section{Evaluation}
\label{sec:evaluation}




    \begin{figure*}[t]
        \centering
        \begin{subfigure}[h]{\columnwidth}
            \centering
            \includegraphics[width=0.9\columnwidth]{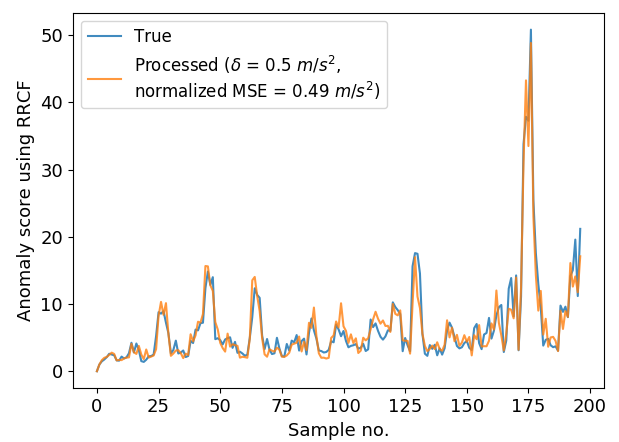}
            \caption[] {Anomaly Detection for error threshold = 0.5 $m/s^2$}    
            \label{fig:anomaly_et_0.5}
        \end{subfigure}
        \hfill
        \begin{subfigure}[h]{\columnwidth}  
            \centering 
            \includegraphics[width=0.9\columnwidth]{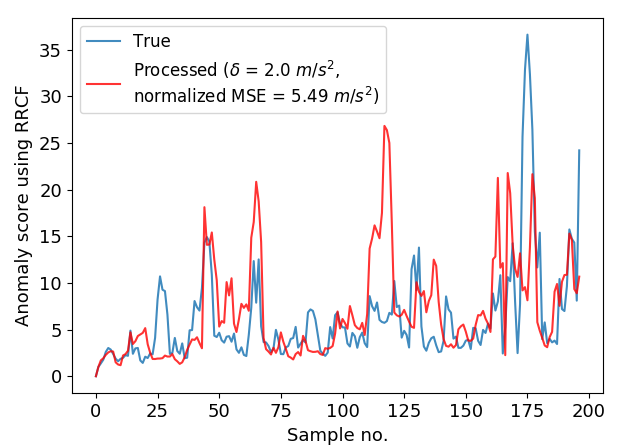}
            \caption[]{Anomaly Detection for error threshold = 2.0 $m/s^2$}
            \label{fig:anomaly_et_2.0}
        \end{subfigure}
        \vskip\baselineskip
        \vspace{-15 pt}
        \caption[]
        {Anomaly Detection for various error thresholds. For threshold up to 1.2, the accuracy of the anomaly detection application is not affected.} 
        \label{fig:anomaly_error_thresh}
        \vspace{-4 mm}
    \end{figure*}

\subsection{Reduction in Data Transfer}

We now present the effectiveness of our approach in reducing the amount of data that needs to be sent from the sensor node to the server. Figure~\ref{fig:true_predicted} shows the transmission reduction for first 200 samples of the acceleration data collected by the sensor node. We evaluated the impact of changing $\delta$ on the percentage of true samples that need to be sent to the server. In Figures~\ref{fig:prediction_0.5} and~\ref{fig:prediction_2.0}, `Processed data' are the samples stored at the server. It includes the true samples (marked in blue) received from the sensor node and the samples predicted (marked in red) by the server when the true sample is not received. For clarity, marking in the figures is done for every other sample.

From Figure~\ref{fig:prediction_0.5}, we can see that even a small $\delta = 0.5$ can reduce the percentage of samples sent to $56.50 \%$, thereby providing a $43.50\%$ reduction in transmission energy consumption. Notably, the processed data is remarkably close to the true data with a normalized Mean Squared Error (MSE) between them = $0.05$. For $\delta$ as high as $2.0$, only $16 \%$ of the data needs to be sent. However, the normalized MSE in this case is very high (= $1.03$) and a substantial distinction can be spotted between the true and processed data (in Figure~\ref{fig:prediction_2.0}). This high an error may not be suitable for most applications and hence it is important to choose an optimal value of the error threshold.

The optimal value of $\delta$ would depend upon a lot of factors such as how sensitive is the desired application to the errors in data samples, how much reduction in the power consumption is required (based on an energy budget) or what is the maximum normalized MSE that can be tolerated. Applications such as anomaly detection are more immune to the errors in data samples and can endure a high $\delta$ value. On the other hand, applications such as sensor tracking of health devices require every individual data sample to be very accurate. Such applications would require $\delta$ to be low. 




\begin{table}
  \centering
  \renewcommand{\arraystretch}{1.2}
  \begin{tabular}{|p{1.2cm}|p{1.4cm}|p{1.4cm}|p{1.4cm}|p{1.4cm}|}
    \hline
    \multirow{2}{2cm}{\textbf{Error Threshold \\ \hspace{0.4cm}($\delta$)}} & \multicolumn{2}{c|}{\textbf{Data sent \%}} & \multicolumn{2}{c|}{\textbf{Normalized MSE}}\\
    \cline{2-5}
    & \textbf{ARIMA Forecasting} & \textbf{Window Forecasting} & \textbf{ARIMA Forecasting} & \textbf{Window Forecasting}\\
    \hline
    0  & 100.00 & 100.00 & 0 & 0 \\ \hline
    0.40  & 61.00 & 66.00 & 0.028 & 0.019  \\ \hline
    0.80 & 41.00 & 41.00 & 0.153 & 0.161 \\ \hline
    1.20 & 26.00 & 32.00 & 0.328 & 0.334  \\ \hline
  \end{tabular}
  \caption{\label{table:window_vs_ARIMA}Data reduction and normalized error comparison between window forecasting and ARIMA (p=3, d=1, q=0) forecasting; number of samples = 200}
\end{table}

\begin{figure}[t]
\vspace{-3mm}
\centering
\includegraphics[width=0.9\columnwidth]{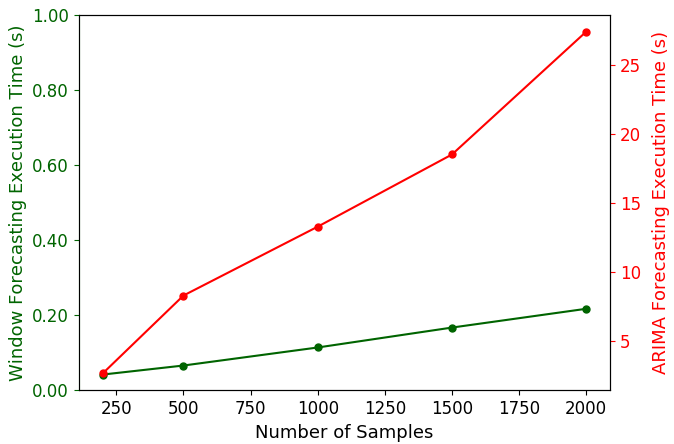}
\caption{\label{fig:window_vs_arima}Comparison between the execution time of \name's window-based and ARIMA forecasting. The slope of the ARIMA plot is $\approx100$X higher than that of ours.}
\centering
\vspace{-4mm}
\end{figure}

\begin{figure*}[t]
        \centering
        \begin{subfigure}[h]{\columnwidth}
            \centering
            \includegraphics[width=0.9\columnwidth]{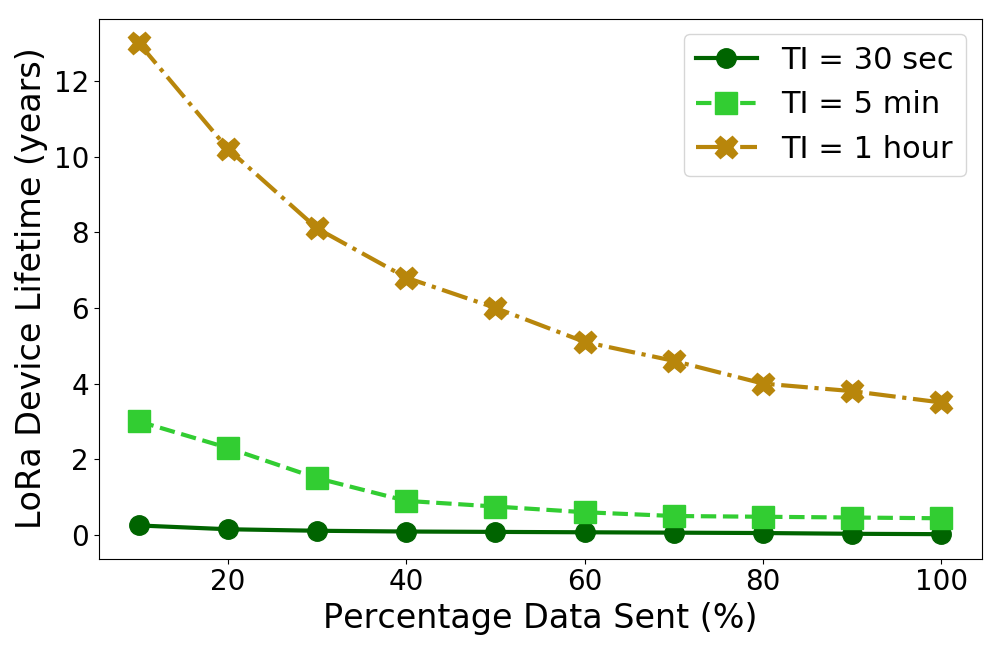}
            \caption[] {Increase in lifetime for LoRa devices}    
            \label{fig:lifetime_increase_LoRa}
        \end{subfigure}
        \hfill
        \begin{subfigure}[h]{\columnwidth}  
            \centering 
            \includegraphics[width=0.9\columnwidth]{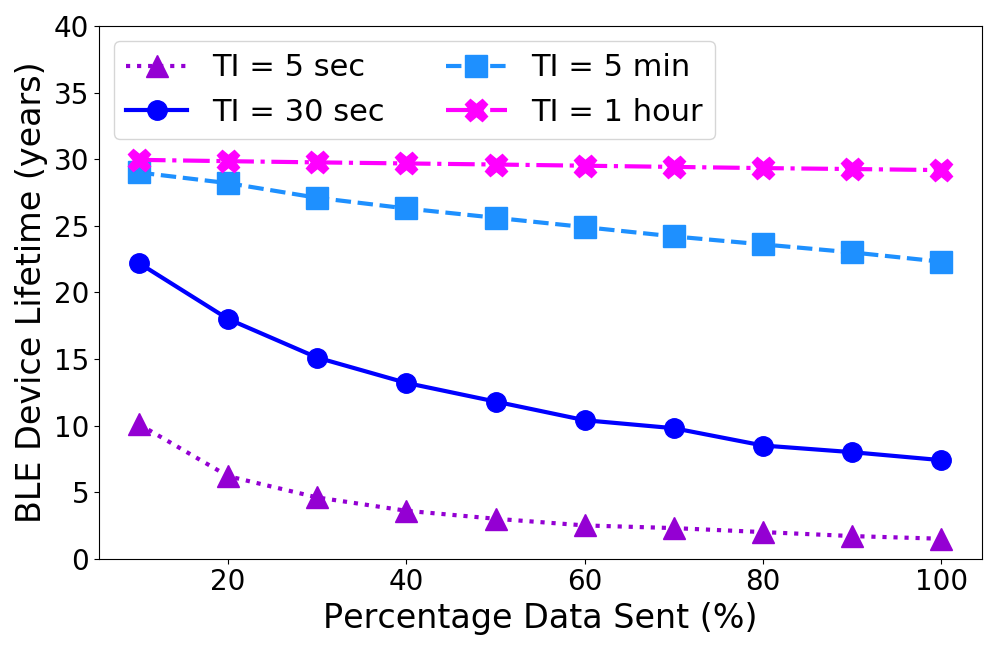}
            \caption[]{Increase in lifetime for BLE devices}
            \label{fig:lifetime_increase_BLE}
        \end{subfigure}
        \vskip\baselineskip
        \vspace{-15 pt}
        \caption[]
        {Increase in IoT sensor node lifetime due to reduction in data transmission; TI is the transmission interval, thus we have low traffic intensity (TI = 1 hour) to high traffic intensity (5 sec in BLE, 30 sec in LoRa). The gain of \name is significantly higher for higher traffic intensity and is available for both wireless technologies LoRa and BLE.}
        \label{fig:lifetime_increase}
        \vspace{-4 mm}
    \end{figure*}

\subsection{Impact on Application: Anomaly Detection}

We have evaluated the impact of reducing data transmission on the accuracy of RRCF anomaly detection application. The aim is to determine if a reduction in the percentage of samples sent negatively impacts the application's performance. Figure~\ref{fig:anomaly_error_thresh} shows a comparison between the anomaly scores of the true data (when all the data samples are sent) and the processed data for different error thresholds ($\delta = 0.5$ for Figure~\ref{fig:anomaly_et_0.5}; $\delta = 2.0$ for Figure~\ref{fig:anomaly_et_2.0}). 

 We are interested in finding out whether the peaks are preserved or not as they are used to detect anomalies. In short, there should not be any false positives (introduction of a false peak) or missed detections (failure to detect a peak) because of the limited data samples sent. We observed that for low values of error threshold, the anomaly score curve for the processed data is comparable to the curve for the true data. For instance, all the anomaly peaks are preserved for $\delta = 0.5$, as shown in Figure~\ref{fig:anomaly_et_0.5}. We observed similar behaviour as $\delta$ increases up to $1.2$, beyond which the correlation between the two curves reduces. Figure~\ref{fig:anomaly_et_2.0} shows the introduction of false positives and missed detection when $\delta = 2.0$. 
 
 It is desirable to have the value of $\delta$ as high as possible (since it reduces the percentage of samples sent), until it starts (negatively) impacting the application's performance. We saw that for anomaly detection, this value is $\approx$ $1.2$, which requires only $32.00\%$ of the data to be sent from the sensor node to the server. It must be noted that if the application's performance is more sensitive to the accuracy of individual data samples, then the desirable $\delta$ for that application would be lower. One such application is displacement computation [not shown due to space constraints].

\subsection{Comparison of Window-Based and ARIMA Forecasting}
\label{subsection:window_vs_ARIMA}

As mentioned in Section~\ref{subsection:window_forecasting}, we use a simple window-based forecasting method for prediction at the sensor nodes and the server. In this section, we compare the performance of our window-based method with the well known ARIMA forecasting technique. In Table~\ref{table:window_vs_ARIMA}, the two forecasting methods are compared based on the percentage of data sent and the normalized MSE for different values of $\delta$. It is evident from the table that the performance of both the methods is similar in terms of the amount of data sent and normalized MSE. This means that both the methods result in a similar increase in the lifetime of the sensor nodes. However, window-based forecasting is computationally much simpler vis-\`a-vis ARIMA, making it more feasible to be implemented on the resource-constrained sensor nodes. 

In Figure~\ref{fig:window_vs_arima}, we compare the execution time of the two methods for different number of data samples. From the figure, we can observe that the execution time for ARIMA is significantly higher than that of the window-based method. Also, the slope of the ARIMA plot is $\approx100$X higher than that of \name's window-based forecasting method. The longer execution time of ARIMA indicates that it cannot sustain a high data rate of incoming samples from the sensor node. Hence, it is preferable to use window-based forecasting method as it provides matched performance with much lower execution time than ARIMA.

\noindent {\bf Reason for using ARIMA as baseline} \\
Our goal is to use lightweight technique so that it can run on an embedded microcontroller. Thus, we do not want to use more complex models like neural network models. We choose ARIMA as this is a superset of many time series models and specific time series models can be derived from it by setting appropriate values for its three parameters. For the baseline evaluation, we optimize the values of the parameters of ARIMA. 

Further, ARIMA despite being an old technique, continues to be successfully applied to time series prediction and is found to be especially suitable for lightweight computation. Publications highlighting this aspect of ARIMA continue to appear till now with regularity~\cite{cook2019anomaly, wang2018short, bhandari2017time}. 

\subsection{Increase in Sensor Node Battery Lifetime}
\label{subsection:lifetime_increase}

In this section, we utilize the technique illustrated in~\cite{comparison_of_device_lifetime} to correlate the reduction in the amount of data transmitted by the sensor node to the increase in the lifetime of the device. We performed the evaluations for LoRa and BLE wireless network technologies. For each wireless technology, we consider varying traffic intensities controlled by the transmission interval (TI), which is the interval between two instances of data transmission. A high value of TI indicates that data is transmitted less frequently thereby corresponding to a low traffic intensity. We consider $2$ standard $AAA$ batteries with initial energy of $13.5$ KJ. Figure~\ref{fig:lifetime_increase} shows the lifetime of the sensor node battery for different values of TI as the percentage of data sent varies ($100\%$ data transmission corresponding to $1$ KB of data).

For LoRa (Figure~\ref{fig:lifetime_increase_LoRa}), if data sent is reduced to $50\%$, the lifetime increases by $71.43\%$ (from $3.5$ years to $6$ years) for low traffic intensity (TI = $1$ hour) and by $300\%$ (from $0.02$ years to $0.08$ years) for high traffic intensity (TI = $30$ sec). LoRa is an expensive technology as long range transmissions consume more energy. If the traffic intensity is high, transmitting $100\%$ of the data would drain the battery very quickly. Therefore, we obtain a higher percentage increase in the lifetime by reducing the data transmission for high traffic intensity. For BLE (Figure~\ref{fig:lifetime_increase_BLE}), if data sent is reduced to $50\%$, the lifetime increases by $1.44\%$ (from $29.18$ years to $29.60$ years) for low traffic intensity (TI = $1$ hour) and by $100\%$ (from $1.5$ years to $3$ years) for high traffic intensity (TI = $5$ sec). BLE does not consume as much power as LoRa, especially if the traffic intensity is low. For low traffic, even with $100\%$ data transmission, the device has a very high battery lifetime of $29.18$ years. Hence, for low traffic, reducing the amount of data sent does not result in a significant percentage increase in the lifetime. However, in many scenarios, BLE is used for streaming data and with a reasonable rate of data, the total data volume can be quite larger and in those cases \name has significant benefit.

\subsection{Second application: Displacement computation for livestock}

We now present the impact of reduction in data transmission on a different application, namely, displacement computation, to track the movement of the animals with ear tags attached. In this application, we used the acceleration data to compute the displacement by double integration. Figure XXX-2 shows a comparison between the displacement computation for the true data (when all the samples are sent) and for the processed data with various error thresholds ($\delta = 0.3, 0.5, \text{and } 1.0 m/s^2$). Unlike the anomaly detection application, this application is more sensitive to the accuracy of individual samples. Therefore, we expect the desirable value of $\delta$ to be lower for this application. The intuition behind this is that the accuracy of this application relies highly upon the numerical values of the samples being accurate compared to the anomaly detection application. In the anomaly detection application, the accuracy depends more on the aberration patterns in the sample values and not much on the numerical values of the samples. 

\begin{figure}
\centering
\includegraphics[width=0.80\textwidth,clip]{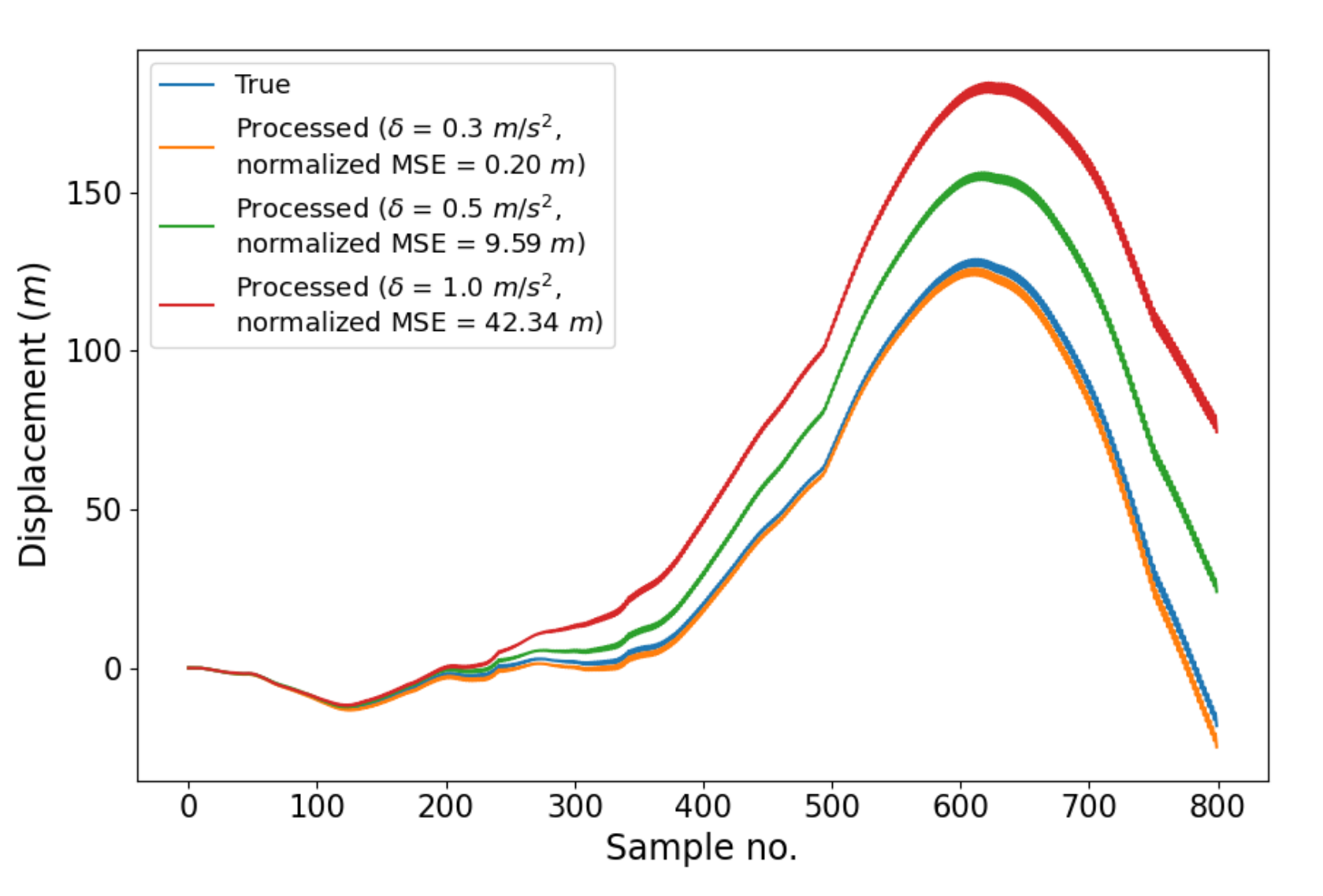}
\caption{Displacement computation for various error thresholds. For threshold up to 0.3 $m/s^2$, the accuracy of displacement computation is not affected.}
\label{fig:displacement-result}
\vspace{-6mm}
\end{figure}

We measured the performance of displacement computation for different $\delta$ values by comparing their displacement curves with the true displacement curve. We quantified the error in terms of normalized MSE. From Figure~\ref{fig:displacement-result}, we can see that the displacement curve for $\delta = 0.3 m/s^2$ is comparable to the true displacement curve and the normalized MSE is very low. For higher values of $\delta$, there is a visible disparity between the curves and the normalized MSE is considerably high. For instance, for $\delta = 1.0 m/s^2$, the normalized MSE is 42.34 m2. Therefore, the desired value of $\delta$ for this application is around $0.3 m/s^2$, which requires 64.62\% of the data to be sent from the sensor node to the server, i.e., 35.38\% lesser data transmission than baseline. Recall that for the anomaly detection application, we could obtain a higher reduction (68\%) in data transmission. This goes on to show that the optimal value of $\delta$ (and hence, the reduction in data transmission obtained) depends highly upon the application under consideration and how sensitive the application quality is to the accuracy of the individual samples.

\section{Conclusion}
\label{sec:conclusion}

Here we presented \name, a lightweight protocol for reducing the amount of data transmissions from IoT sensors to the server resulting in an increase in the battery lifetime of the sensor nodes. We introduced a window-based time series forecasting mechanism that forms a key element of our data reduction protocol and can be easily implemented on even the most resource constrained sensor nodes, such as ear tags put on livestock. Compared to the state-of-the-art ARIMA forecasting, \name is considerably faster ($99 \%$ lower execution time) while providing similar data reduction. We evaluated \name on different wireless network technologies such as LoRa and BLE and obtained more than $60\%$ reduction in data transmission (almost $2$X increase in sensor node battery lifetime). We identified the correlation between the data reduction and error tolerance of different applications and provide a configurable error threshold to ensure that the accuracy of the end applications is not impacted by the reduction in data transmissions.

\small
\setlength{\bibsep}{0.5pt}
\bibliography{References}

\end{document}